\documentclass[12pt]{article}
\title{Series of zeta values, the Stieltjes constants, and a sum $S_\gamma(n)$}
\author{Mark W. Coffey\\
Department of Physics\\
Colorado School of Mines\\
Golden, CO  80401\\
(Received $\mbox{~~~~~~~~~~~~~~~~~~~~~~~~~~~~~~~2009}$)}
\date{February 24, 2009}
\begin{document}
\maketitle
\baselineskip=25 pt
\begin{abstract}

We present a variety of series representations of the Stieltjes and related
constants, the Stieltjes constants being the coefficients of the Laurent
expansion of the Hurwitz zeta function $\zeta(s,a)$ about $s=1$. 
Additionally we obtain series and integral representations of a sum
$S_\gamma(n)$ formed as an alternating binomial series from the Stieltjes
constants.  The slowly varying sum $S_\gamma(n)+n$ is an important subsum
in application of the Li criterion for the Riemann hypothesis.

\end{abstract}
 
\vspace{.25cm}
\baselineskip=15pt
\centerline{\bf Key words and phrases}
\medskip 

\noindent
Hurwitz zeta function, Riemann zeta function, digamma function, polygamma function, 
series representation, integral representation, Stieltjes constants, polylogarithms,
Laguerre polynomials, Dirichlet $L$ function

\vfill
\centerline{\bf 2000 AMS subject classification}

11M06, 11M35, 33B15
 
\baselineskip=25pt
\pagebreak
\centerline{\bf Introduction}

The Hurwitz zeta function, defined by $\zeta(s,a)=\sum_{n=0}^\infty (n+a)^{-s}$
for Re $s>1$ and Re $a>0$, extends to a meromorphic function in the entire
complex $s$-plane.  This analytic continuation to C has a simple pole of
residue one.  This is reflected in the Laurent expansion 
$$\zeta(s,a)={1 \over {s-1}}+ \sum_{n=0}^\infty {{(-1)^n} \over {n!}}\gamma_n(a)
(s-1)^n, \eqno(1)$$
wherein $\gamma_k(a)$ are designated the Stieltjes constants
\cite{coffeyjmaa,coffeyprsa,ivic,liang,stieltjes,wilton2} and $\gamma_0(a)=-\psi
(a)$, where $\psi=\Gamma'/\Gamma$ is the digamma function.
The constants $\gamma_k(a)$ can be written in the form
$$\gamma_k(a)=\lim_{N \to \infty}\left (\sum_{m=0}^N {1 \over {m+a}}\ln^k (m+a)-
{{\ln^{k+1} {(N+a)}} \over {k+1}}\right ). \eqno(2)$$
In certain special cases including $a \geq 1$ a positive integer and $a=1/2$
the Hurwitz zeta function may be expressed in terms of the Riemann zeta function
$\zeta(s)$. 
In the case of $a=1$, one puts $\gamma_k(1)=\gamma_k$ by convention.

In this paper we present an array of series representations of the Stieltjes
and related constants.  Moreover we obtain series and integral representations 
of the sum
$$S_\gamma(n) \equiv \sum_{k=1}^n {{(-1)^k} \over {(k-1)!}} {n \choose k}
\gamma_{k-1}.  \eqno(3)$$
Putting $P_1(t)=B_1(t-[t])=t-[t]-1/2$ the first periodized
Bernoulli polynomial (e.g., \cite{ivic,titch}) we may also write  
$$S_\gamma(n)=\int_1^\infty {1 \over t} L_{n-1}^1(\ln t) dP_1(t), \eqno(4)$$  
where $L_n^\alpha$ is the Laguerre polynomial of degree $n$.
We have shown elsewhere \cite{inprep} that $S_\gamma(n)$ is a component
sum of the sum $S_2(n)$ arising in application of the Li criterion for the
Riemann hypothesis \cite{coffeympag,coffey08}.  We have demonstrated \cite{inprep}
that $S_\gamma(n)+n=O(n^{1/4})$.  Whereas the Stieltjes constants
are very closely related to derivatives of the zeta function at $s=1$, we
develop a series representation for them in terms of zeta function 
derivatives at integers $j \geq 2$.

For a collection of
formulae involving series of zeta function values we mention Ref. \cite{sri}.

We introduce the constants
$$d_j(a) \equiv \int_0^\infty {{\ln^{j-1} (t+a)} \over {(t+a)}} \ln\left({{[t]+a+1}
\over {t+a+1}}\right )dt, ~~~~~~~~~~j \geq 1,\eqno(5)$$
and
$$c_j \equiv \int_1^\infty {{\ln^{j-1} t} \over t} \ln\left({{[t]+1}
\over {t+1}}\right )dt=d_j(1), ~~~~~~~~~~j \geq 1.  \eqno(6)$$
The approximate values of the first few of the latter constants are
$c_1 \simeq -0.334$, $c_2 \simeq -0.433$, and $c_3 \simeq -0.93$.
We have in general from the definition (6) that $c_j < 0$.  We then have 
{\newline \bf Proposition 1}.  For integers $j \geq 1$ we have (a)
$$\gamma_j = (-1)^j\sum_{m=2}^\infty {{(-1)^m} \over m} \zeta^{(j)}(m)
-(-1)^j j! (1-2^{-j})\zeta(j+1)- j c_j,  \eqno(7)$$
(b)
$$\gamma_j(a) = (-1)^j\sum_{m=2}^\infty {{(-1)^m} \over m} \zeta^{(j)}(m,a)
-j\sum_{\ell=0}^{j-1} (-1)^\ell \ell !{{j-1} \choose \ell} \ln^{j-\ell-1} a ~\mbox{Li}_{\ell+2}(-a)$$
$$-jd_j(a)-\ln^{j+1} a, \eqno(8)$$
(c) for integers $n \geq 2$
$$S_\gamma(n)=1-\ln 2-(2 \ln 2+\gamma-1)n+\int_1^\infty {{P_1(x)} \over x}
\left[{1 \over x}-\ln\left(1+{1 \over x}\right)\right] L_{n-2}^2(\ln x)dx$$
$$-\int_1^\infty {{P_1(x)} \over {x^2(1+x)}}\left[n-L_{n-1}^1(\ln x)\right]dx
+\sum_{m=3}^\infty {{(-1)^m} \over m}\left({{m-2} \over {m-1}}\right)^n$$
$$+\int_0^\infty \left[L_{n-1}^1(-x)-L_{n-1}^1(-x/2)\right]{{dx}
\over {(e^x -1)}}+\int_1^\infty {{L_{n-2}^2(\ln t)} \over t}
\ln\left({{[t]+1} \over {t+1}}\right )dt,  \eqno(9)$$
and (d) with $f(t)=\ln(t+1)/t(t+1)$
$$\sum_{j=1}^\infty {{(-1)^{j-1}} \over {j!}}(jc_j)={\pi^2 \over {12}}
+{1 \over 2}\ln^2 2-{7 \over 4}\ln 2+\int_1^\infty P_1(t)f'(t)dt,  \eqno(10a)$$
while with $g(t,a)=\ln(t+a)/(t+a-1)(t+a)$
$$\sum_{j=1}^\infty {{(-1)^{j-1}} \over {j!}}(jd_j)=\sum_{k=1}^\infty g(k,a)
-\left(1+{1 \over a}\right)\ln(a+1)-\ln a$$
$$=\zeta(2)+{1 \over 2}\ln^2 (a+1)+\mbox{Li}_2(-a)+{1 \over 2}{{\ln(a+1)} \over
{a(a+1)}}-\left(1+{1 \over a}\right)\ln(a+1)-\ln a+\int_1^\infty P_1(t)g'(t,a)dt.
\eqno(10b)$$
In Eq. (8) Li$_j$ is the polylogarithm function and in Eq. (9) $\gamma$ is the 
Euler constant.

After the proof of this Proposition we present additional summatory results
and a generalization of them.  These other Propositions pertain to $\gamma_0(a)$
and sums of Hurwitz zeta function values.  The final section and Proposition       
treat the coefficients of the Laurent expansion of the logarithmic derivative
of the Riemann zeta function about $s=1$.

\pagebreak
\centerline{\bf Proof of Proposition 1}
\medskip

For part (a) we begin by forming
$$\sum_{m=2}^\infty {{(-1)^m} \over m}(-1)^j \zeta^{(j)}(m)
=\sum_{k=2}^\infty \sum_{m=2}^\infty {{(-1)^m} \over {mk^m}} \ln^j k$$
$$=\sum_{k=2}^\infty \ln^j k\left[{1 \over k} -\ln\left({{k+1} \over k}\right)
\right]$$
$$=\lim_{x \to \infty}\left \{\sum_{k=2}^x {{\ln^j k} \over k}+\sum_{k=2}^x
[\ln k-\ln(k+1)]\ln^j k\right \}.  \eqno(11)$$
We then apply summation by parts to write
$$\sum_{k=2}^x [\ln k-\ln(k+1)]\ln^j k=-\ln([x]+1)\ln^j x+j\int_1^x \ln([t]+1)
{{\ln^{j-1} t} \over t}dt$$
$$=-\ln([x]+1)\ln^j x+j\left(\int_1^x \ln(t+1){{\ln^{j-1} t} \over t}dt+c_j^{(x)}
\right),  \eqno(12)$$
where
$$c_j^{(x)} \equiv \int_1^x {{\ln^{j-1} t} \over t} \ln\left({{[t]+1} \over
{t+1}}\right )dt, ~~~~~~~~~~j \geq 1.  \eqno(13)$$
The integral on the right side of Eq. (12) is evaluated in terms of polylogarithms
Li$_j(z)=\sum_{k=1}^\infty z^k/k^j$ for $|z| \leq 1$ and otherwise analytically
continued, by making use of the following.

{\bf Lemma 1}.  We have
$$\int_1^x \ln(t+1){{\ln^{j-1} t} \over t}dt=\sum_{\ell=0}^{j-1}(-1)^{\ell-1}
\ell! {{j-1} \choose \ell}\ln^{j-\ell-1} x ~\mbox{Li}_{\ell+2}(-x)+(-1)^j(j-1)!
(1-2^{-j})\zeta(j+1).  \eqno(14)$$
To prove the Lemma, we make use of the representation \cite{grad} (p. 568)
$$\int_0^1 \ln^r x \ln(1-bx) {{dx} \over x}=(-1)^{r-1} \Gamma(r+1)\mbox{Li}_{r+2}
(b), ~~~~~~~~\mbox{Re} ~r >-1,\eqno(15)$$
where $\Gamma$ is the Gamma function, and the relation \cite{sri}
$$\mbox{Li}_{j+1}(-1)=-(1-2^{-j})\zeta(j+1).  \eqno(16)$$
We write
$$\int_1^x \ln(t+1){{\ln^{j-1} t} \over t}dt=\int_0^x \ln(t+1){{\ln^{j-1} t} 
\over t}dt -\int_0^1 \ln(t+1){{\ln^{j-1} t} \over t}dt$$
$$=\int_0^1 \ln(1+xw)(\ln x+\ln w)^{j-1} {{dw} \over w}-(-1)^j(j-1)!
\mbox{Li}_{j+1}(-1).    \eqno(17)$$
We then binomially expand the integrand in this equation and apply both
Eqs. (15) and (16) and the Lemma follows.

From Lemma 1 we may now write for Eq. (12)
$$\sum_{k=2}^x [\ln k-\ln(k+1)]\ln^j k=-\ln([x]+1)\ln^j x$$
$$+j\sum_{\ell=0}^{j-1}
(-1)^{\ell-1}\ell! {{j-1} \choose \ell}\ln^{j-\ell-1} x \mbox{Li}_{\ell+2}(-x)
+(-1)^jj!(1-2^{-j})\zeta(j+1)+jc_j(x).  \eqno(18)$$
From the relation between Li$_n(-x)$ and Li$_n(-1/x)$ (e.g., \cite{sri}, p.
116)
$$\mbox{Li}_n(-x) +(-1)^n\mbox{Li}_n\left(-{1 \over x}\right)=-{1 \over {n!}}
\ln^n x+2\sum_{k=1}^{[n/2]} {{\ln^{n-2k} x} \over {(n-2k)!}} \mbox{Li}_{2k}
(-1),  \eqno(19)$$
we may read off the asymptotic form  
$$\mbox{Li}_n(-x) \sim -{1 \over {n!}}
\ln^n x+2\sum_{k=1}^{[n/2]} {{\ln^{n-2k} x} \over {(n-2k)!}} \mbox{Li}_{2k}
(-1), ~~~~~~~~x \to \infty.  \eqno(20)$$
Then inserting this expression into the sum on the right side of Eq. (17) gives
$$j\sum_{\ell=0}^{j-1} (-1)^{\ell-1}\ell! {{j-1} \choose \ell}\ln^{j-\ell-1} x ~
\mbox{Li}_{\ell+2}(-x)$$
$$\sim j\sum_{\ell=0}^{j-1} (-1)^{\ell-1}\left[-{{\ln^{j+1} x}
\over {(\ell+1)(\ell+2)}}+2\sum_{k=1}^{[\ell+2]/2} {{\ln^{\ell+2-2k} x} \over
{(\ell+2-2k)!}}\mbox{Li}_{2k}(-1)\right]$$
$$={j \over {j+1}}\ln^{j+1} x$$
$$+2j\ln^{j+1}x\sum_{\ell=0}^{j-1} (-1)^{\ell-1}
{{j-1} \choose \ell} \sum_{k=1}^{[\ell+2]/2} {{\ell!} \over {(\ell+2-2k)!}}
\ln^{-2k} x \mbox{Li}_{2k}(-1).  \eqno(21)$$
For a given $k$, the sum over $\ell$ on the right side of Eq. (21) gives
zero as
$$\sum_{\ell=0}^{j-1} (-1)^{\ell-1}{{j-1} \choose \ell}{{\ell!} \over
{(\ell+2-2k)!}}=-{1 \over {(j-2k+1)}}{1 \over {\Gamma(2-2k)}}$$
$$=-{1 \over {(j-2k+1})\pi}\Gamma(2k-1)\sin \pi(2k-1).  \eqno(22)$$

We now have from Eqs. (11), (12), and (21) that
$$\sum_{m=2}^\infty {{(-1)^m} \over m}(-1)^j \zeta^{(j)}(m)
=\lim_{x \to \infty}\left\{\sum_{k=2}^x {{\ln^j k} \over k}-\ln^{j+1} x
+O\left({{\ln^j x} \over x}\right)\right.$$
$$\left.+{j \over {j+1}}\ln^{j+1} x +j c_j(x)
+(-1)^j j!(1-2^{-j})\zeta(j+1)\right\}$$
$$=\lim_{x \to \infty} \left(\sum_{k=2}^x {{\ln^j k} \over k}-{{\ln^{j+1} x}
\over {j+1}}\right)+jc_j+(-1)^j j!(1-2^{-j})\zeta(j+1)$$
$$=\gamma_j +jc_j+(-1)^j j!(1-2^{-j})\zeta(j+1).  \eqno(23)$$ 
In the last step we used relation (2).  For the next to last step, since we
have 
$$\ln\left({{[t]+1} \over {t+1}}\right )=O\left({{\{t\}} \over t}\right),
\eqno(24)$$
where $\{t\}$ denotes the fractional part of $t$, the constants $c_j$ are
certainly well defined and $\lim_{x \to \infty} c_j^{(x)} = c_j$.
We have demonstrated part (a).

For part (b) we proceed similarly to part (a).  We first have
$$\sum_{m=2}^\infty {{(-1)^m} \over m}(-1)^j \zeta^{(j)}(m,a)
=\lim_{x \to \infty}\left \{\sum_{k=0}^x {{\ln^j (k+a)} \over {k+a}}
+\sum_{k=0}^x [\ln (k+a)-\ln(k+a+1)]\ln^j (k+a)\right \}.  \eqno(25)$$
We next apply summation by parts, writing
$$\sum_{k=0}^x [\ln (k+a)-\ln(k+a+1)]\ln^j (k+a)=[\ln a-\ln(a+1)]\ln^j a
+[\ln(a+1)-\ln([x]+a+1)]\ln^j(x+a)$$
$$-j\int_0^x[\ln(a+1)-\ln([t]+a+1)]{{\ln^{j-1} (t+a)} \over {t+a}}dt.  \eqno(26)$$
The last term here can be written
$$-j\int_0^x[\ln(a+1)-\ln([t]+a+1)]{{\ln^{j-1} (t+a)} \over {t+a}}dt
=\ln(a+1)[\ln^j a-\ln^j(x+a)]$$
$$+j\left(\int_0^x\ln(t+a+1){{\ln^{j-1} (t+a)} \over {t+a}}dt+d_j^{(x)}(a)\right), \eqno(27)$$
where we have put
$$d_j^{(x)}(a) \equiv \int_0^x {{\ln^{j-1} (t+a)} \over {(t+a)}} \ln\left({{[t]+a+1}\over {t+a+1}}\right )dt, ~~~~~~~~~~j \geq 1.\eqno(28)$$

The integral of Eq. (27) is obtained from Lemma 1 as
$$\int_0^x \ln(t+a+1){{\ln^{j-1} (t+a)} \over {(t+a)}}dt
=\left(\int_1^{x+a}-\int_1^a\right)\ln(u+1) {{\ln^{j-1} u} \over u}du$$
$$=\sum_{\ell=0}^{j-1}
(-1)^{\ell-1} \ell! {{j-1} \choose \ell}\left[\ln^{j-\ell-1} (x+a) ~\mbox{Li}_{\ell+2}
(-x-a)
-\ln^{j-\ell-1} a ~\mbox{Li}_{\ell+2}(-a)\right].  \eqno(29)$$
We combine Eqs. (26), (27) and (29), giving
$$\sum_{k=0}^x [\ln (k+a)-\ln(k+a+1)]\ln^j (k+a)=[\ln a-\ln(a+1)]\ln^j a
+[\ln(a+1)-\ln([x]+a+1)]\ln^j(x+a)$$
$$+\ln(a+1)[\ln^j a-\ln^j(x+a)]+jd_j^{(x)}(a)$$
$$+j\sum_{\ell=0}^{j-1}(-1)^{\ell-1} \ell! {{j-1} \choose \ell}\left[\ln^{j-\ell-1} (x+a) ~\mbox{Li}_{\ell+2}(-x-a)-\ln^{j-\ell-1} a ~\mbox{Li}_{\ell+2}(-a)\right].  \eqno(30)$$

We then make use of the asymptotic relation (20), obtaining
$$\sum_{m=2}^\infty {{(-1)^m} \over m}(-1)^j \zeta^{(j)}(m,a)
=\lim_{x \to \infty}\left\{\sum_{k=0}^x {{\ln^j (k+a)} \over {k+a}}+\ln^{j+1} a
-\ln^{j+1} (x+a)+O\left({{\ln^j (x+a)} \over {x+a}}\right)\right.$$
$$\left. +{j \over {j+1}}\ln^{j+1} (x+a) +j d_j^{(x)}(a)
+j\sum_{\ell=0}^{j-1} (-1)^{\ell}\ell !{{j-1} \choose \ell} \ln^{j-\ell-1} a ~\mbox{Li}_{\ell+2}(-a)\right\}$$
$$=\lim_{x \to \infty} \left(\sum_{k=0}^x {{\ln^j (k+a)} \over {(k+a)}}
-{{\ln^{j+1} (x+a)}\over {j+1}}\right)+\ln^{j+1} a + jd_j(a)
+j\sum_{\ell=0}^{j-1} (-1)^{\ell}\ell !{{j-1} \choose \ell} \ln^{j-\ell-1} a ~\mbox{Li}_{\ell+2}(-a)$$
$$=\gamma_j(a)+\ln^{j+1}a +jd_j(a)+j\sum_{\ell=0}^{j-1} (-1)^{\ell}\ell !{{j-1} 
\choose \ell} \ln^{j-\ell-1} a ~\mbox{Li}_{\ell+2}(-a).  \eqno(31)$$ 
Part (b) has been demonstrated.

Remark.  When $a=1$, the only term remaining in the sum on the right side of Eq. (8)
is $\ell=j-1$.  By relation (16), we then see the proper reduction of part (b) to
part (a). 

For the proof of part (c) we make repeated use of 
{\newline \bf Lemma 2}.  We have
$$\sum_{j=\nu}^n {{(-1)^{j-1}} \over {(j-\nu)!}}{n \choose j} w^{j-\nu}
=(-1)^{\nu-1} L_{n-\nu}^\nu (w),  \eqno(32)$$
that follows from the power series form of the Laguerre polynomials.
We also apply
{\newline \bf Lemma 3}.  We have for integers $n \geq 1$
$$\zeta^{(n)}(s)=(-1)^n n\int_1^\infty {{P_1(x)} \over x^{s+1}}\ln^{n-1} x ~dx
-(-1)^n s\int_1^\infty {{P_1(x)} \over x^{s+1}}\ln^n x ~dx +{{(-1)^n n!} \over
{(s-1)^{n+1}}}.  \eqno(33)$$
Lemma 3 follows from induction on the integral representation \cite{titch} (p. 14)
$$\zeta(s)={1 \over 2}+{1 \over {s-1}}-s\int_1^\infty {{P_1(x)} \over x^{s+1}}dx.
\eqno(34)$$
We combine the definition (3) of $S_\gamma(n)$, Lemma 2, the definition (6)
of $c_j$, and the expression (7) for $\gamma_j$.  For the $\zeta(j)$ values in
Eq. (7) we apply the integral representation for Re $s>1$ and Re $a>0$
$$\Gamma(s)\zeta(s,a)=\int_0^\infty {{t^{s-1} e^{-(a-1)t}} \over 
{e^t-1}}dt,  \eqno(35)$$
at $a=1$.  We then obtain
$$S_\gamma(n)=-\gamma n-\sum_{k=2}^n {1 \over {(k-1)!}}{n \choose k}
\sum_{m=2}^\infty {{(-1)^m} \over m}\zeta^{(k-1)}(m)$$
$$+\int_0^\infty \left[L_{n-1}^1(-x)-L_{n-1}^1(-x/2)\right]{{dx}
\over {(e^x -1)}}+\int_1^\infty {{L_{n-2}^2(\ln t)} \over t}
\ln\left({{[t]+1} \over {t+1}}\right )dt.  \eqno(36)$$

We use Lemma 3 to re-express the double series on the right side of Eq. (36):
$$\sum_{k=2}^n {1 \over {(k-1)!}}{n \choose k} \sum_{m=2}^\infty {{(-1)^m} \over m}
\zeta^{(k-1)}(m)=\sum_{m=2}^\infty {{(-1)^m} \over m}\sum_{k=2}^n (-1)^{k-1}
{n \choose k}\left[{1 \over {(k-2)!}}\int_1^\infty {{P_1(x)} \over x^{m+1}}\ln^{k-2}
x ~dx\right.$$
$$\left.+{1 \over {(m-1)^k}}-{m \over {(k-1)!}}\int_1^\infty {{P_1(x)} \over x^{m+1}}
\ln^{k-1}x ~dx\right]$$
$$=\sum_{m=2}^\infty {{(-1)^m} \over m}\left[-\int_1^\infty {{P_1(x)} \over x^{m+1}}
L_{n-2}^2(\ln x)dx+1-{n \over {(m-1)}}-\left({{m-2} \over {m-1}}\right)^n \right.$$
$$\left.+ m\int_1^\infty {{P_1(x)} \over x^{m+1}}[n-L_{n-1}^1(\ln x)] dx \right]. 
\eqno(37)$$
We then make use of the elementary series
$$\sum_{m=2}^\infty {{(-1)^m} \over m}=1-\ln 2, ~~~~~~~~~~~
-n\sum_{m=2}^\infty {{(-1)^m} \over {m(m-1)}}=n-2n \ln 2,  \eqno(38a)$$
$$\sum_{m=2}^\infty {{(-1)^m} \over m}{1 \over x^{m+1}}={1 \over x^2}-{1 \over x}
\ln\left(1+{1 \over x}\right), ~~~~~~~~~\sum_{m=2}^\infty {{(-1)^m} \over x^{m+1}}
={1 \over {x^2(1+x)}},  \eqno(38b)$$
and part (c) of Proposition 1 follows.

For part (d) we may first form using Eq. (6)
$$\sum_{j=1}^\infty {{u^{j-1}} \over {j!}}(jc_j)=\int_1^\infty t^{u-1} 
\ln\left({{[t]+1} \over {t+1}}\right )dt, ~~~~~~~~~~\mbox{Re}~u \leq 0.
\eqno(39)$$
We may note in passing that this equation properly recovers the expression
for $c_1$ at $u=0$.  At $u=-1$ we have
$$\sum_{j=1}^\infty {{(-1)^{j-1}} \over {j!}}(jc_j)=\sum_{k=1}^\infty
\int_k^{k+1} {{\ln(k+1)} \over t^2} dt - \int_1^\infty {{\ln(t+1)} \over t^2}
dt.  \eqno(40)$$
An integration by parts gives the integral
$$\int_1^x {{\ln(t+1)} \over t^2}dt=\ln x -\left(1+{1 \over x}\right)\ln(x+1)
+2\ln 2.  \eqno(41)$$
Therefore we have from Eq. (40)
$$\sum_{j=1}^\infty {{(-1)^{j-1}} \over {j!}}(jc_j)=\sum_{k=1}^\infty {{\ln(k+1)}
\over {k(k+1)}} - 2 \ln 2.  \eqno(42)$$
We next apply Euler-Maclaurin summation to the sum on the right side of Eq. (36):
$$\sum_{k=1}^\infty {{\ln(k+1)} \over {k(k+1)}}=\int_1^\infty f(t) dt+{1 \over 2}
f(1)+\int_1^\infty P_1(t)f'(t)dt$$
$$={\pi^2 \over {12}}+{1 \over 2}\ln^2 2+{1 \over 4} \ln 2
+\int_1^\infty P_1(t)f'(t)dt.  \eqno(43)$$
Combining Eqs. (42) and (43) gives Eq. (10a).  

Very similarly, we have
$$\sum_{j=1}^\infty {{u^{j-1}} \over {j!}}(jd_j)=\int_0^\infty (t+a)^{u-1} 
\ln\left({{[t]+a+1} \over {t+a+1}}\right )dt,$$
giving
$$\sum_{j=1}^\infty {{(-1)^{j-1}} \over {j!}}(jd_j)=\sum_{k=1}^\infty g(k,a)
-\left(1+{1 \over a}\right)\ln(a+1)-\ln a.$$
Applying Euler-Maclaurin summation then gives Eq. (10b), completing part (d) and
the Proposition.

Remarks. (i) Concerning approximate numerical values in part (d) we have
$$\sum_{k=1}^\infty {{\ln(k+1)} \over {k(k+1)}} \simeq 1.25774688694, ~~~~~~~~~
\int_1^\infty P_1(t)f'(t)dt \simeq 0.0233638, \eqno(44a)$$
and
$$\sum_{j=1}^\infty {{(-1)^{j-1}} \over {j!}}(jc_j) \simeq -0.12816.  \eqno(44b)$$
(ii) The interesting sum of Eq. (44a) has a multitude of equivalent expressions,
including a great variety of series and integral representations.  By splitting
the sum over even and odd contributions we have
$$\sum_{k=1}^\infty {{\ln(k+1)} \over {k(k+1)}} =\ln^2 2 +{1 \over 2}
\sum_{k=1}^\infty {1 \over {(2k+1)}}\left[{{\ln(k+1)} \over {k+1}}+{{\ln(2k+1)} 
\over k}\right].  \eqno(45)$$

By using an integral representation for $\ln z$, we have derived the 
representations
$$\sum_{k=1}^\infty {{\ln(k+1)} \over {k(k+1)}} =\int_0^\infty {{(e^{-t}-1)} 
\over t}\ln(1-e^{-t}) dt
=\int_0^1 {{(1-x)\ln(1-x)} \over {x \ln x}}dx\eqno(46a)$$
$$=\int_0^\infty e^{-t}[1+\ln(1-e^{-t})]\ln t ~dt=-\gamma+\int_0^\infty e^{-t} \ln t
\ln(1-e^{-t}) dt,  \eqno(46b)$$
$$=-\gamma+\int_0^1 e^{-t} \ln t\ln(1-e^{-t}) dt-\sum_{k=1}^\infty {1 \over k}
{{\Gamma(0,k+1)} \over {(k+1)}},  \eqno(46c)$$
where the second line follows from the first by integration by parts and 
$\Gamma(x,y)$ is the incomplete Gamma function.  By using the Maclaurin series
for $\ln(1-z)$ in either integral of Eq. (46a) we find
$$\sum_{k=1}^\infty {{\ln(k+1)} \over {k(k+1)}} = \sum_{r=1}^\infty {1 \over r}
\ln\left(1+ {1 \over r}\right).  \eqno(46d)$$
In addition, we have by partial summation
$$\sum_{k=1}^\infty {{\ln(k+1)} \over {k(k+1)}}=-\sum_{n=1}^\infty H_n\left[
{{\ln(n+2)} \over {n+2}}-{{\ln(n+1)} \over {n+1}}\right], \eqno(46e)$$ 
where $H_n \equiv \sum_{k=1}^n 1/k$ are the harmonic numbers, and we next show how
this expression may be alternatively obtained from the integral representation (46b).
We use the generating function 
$$-\ln(1-z)=(1-z)\sum_{n=1}^\infty H_n z^n \eqno(47)$$
at $z=e^{-t}$.  We substitute for $\ln(1-e^{-t})$ on the right side of Eq. (46b),
interchange summation and integration, justified by the absolute convergence of
the integral, and find that
$$\int_0^\infty e^{-t} \ln t \ln(1-e^{-t}) dt=-\sum_{n=1}^\infty H_n\int_0^\infty
[e^{-(n+1)t}-e^{-(n+2)t}]\ln t ~dt \eqno(48a)$$
$$=\sum_{n=1}^\infty H_n\left[\gamma\left({1 \over {n+1}}-{1 \over {n+2}}\right)
+{{\ln(n+1)} \over {n+1}}-{{\ln(n+2)} \over {n+2}}\right]. \eqno(48b)$$
We note that simply $H_{n-1}=H_n-1/n$ and recover Eq. (46e).

Using the generating function (47) and the right-most integral of Eq. 
(46a) we find that 
$$\sum_{k=1}^\infty {{\ln(k+1)} \over {k(k+1)}}=\sum_{n=1}^\infty H_n
\int_0^1 {{(1-x)^2 x^{n-1}} \over {\ln x}} dx
=\sum_{n=1}^\infty H_n[\ln n -2\ln(n+1)+\ln(n+2)].  \eqno(49)$$
The integral of Eq. (49) may be found from the special case of the Beta function
$B(x,y)$ integral
$$\int_0^1 (1-x)^2 x^{q-1}dx=B(3,q)={2 \over {q(q+1)(q+2)}}={1 \over q}
-{2 \over {q+1}}+{1 \over {q+2}},  \eqno(50)$$
by integrating over $q$.  Similarly we may determine Eq. (43) by using the 
other integral of Eq. (46a).

Other forms of the sum of Eq. (44a) may be obtained from Eq. (46b)
and the generating function
$$g_j(z) \equiv \sum_{n=1}^\infty n^j (H_n-1)z^n, ~~~~~~~~~|z| < 1.  \eqno(51)$$
We may advance to $g_{j+1}$ from $g_j$ with the use of the operator $\theta
\equiv z (d/dz)$:  $g_{j+1}(z) = \theta g_j(z)$.
We have $g_1(z)=-z \ln(1-z)/(1-z)^2$ and obtain
$$\sum_{k=1}^\infty {{\ln(k+1)} \over {k(k+1)}}=\sum_{n=1}^\infty n(H_n-1)\left[
{{\ln n} \over n}-2{{\ln(n+1)} \over {n+1}}+{{\ln(n+2)} \over {n+2}}\right]. \eqno(52)$$    
One could pose the question whether all such expressions obtainable from the use of 
$g_j(z)$ are also obtainable through partial summation.  

Finally, we illustrate the application of a Binet formula for the digamma
function \cite{grad} (p. 943), that we use as a representation of $\ln(k+1)$.
We have 
$$\sum_{k=1}^\infty {{\ln(k+1)} \over {k(k+1)}}=\sum_{k=1}^\infty{{\psi(k+1)}
\over {k(k+1)}}+{1 \over 2}\sum_{k=1}^\infty{1 \over {k(k+1)^2}}+2\sum_{k=1}^
\infty \int_0^\infty {t \over {[t^2+(k+1)^2]}}{{dt} \over {(e^{2\pi t}-1)}}.
\eqno(53)$$
We evaluate the $\psi$ sum of this equation by inserting an integral
representation for it and interchanging operations.  For the sum on the right
side of Eq. (53) we apply the partial fraction decomposition for the hyperbolic
cotangent function \cite{grad} (p. 36).  We obtain
$$\sum_{k=1}^\infty {{\ln(k+1)} \over {k(k+1)}}=1-\gamma+{\pi^2 \over {12}}
+2\int_0^\infty \left[{\pi \over {2t}}\coth \pi t-{1 \over {2t^2}}-{1 \over 
{t^2+1}}\right]{{t dt} \over {(e^{2\pi t}-1)}} \eqno(54a)$$
$$={3 \over 2}-2\gamma+{\pi^2 \over {12}}+\int_0^\infty\left(\pi \coth \pi t
-{1 \over t}\right){{dt} \over {(e^{2\pi t}-1)}}.  \eqno(54b)$$
For the last step we have used \cite{grad} (p. 328).

(iii) The alternating sum on the right side of Eq. (9),
$$\sum_{m=3}^\infty {{(-1)^m} \over m}\left({{m-2} \over {m-1}}\right)^n
\simeq \sum_{m=3}^\infty {{(-1)^m} \over m} e^{-n/(m-1)}, \eqno(55)$$ 
is exponentially decreasing with $n$.  The last integral term on the right side
of Eq. (9) is over estimated by $n-1$.

\medskip
\centerline{\bf A Corollary and further remarks}
\medskip

The method of Proposition 1(b) also has relevance to Dirichlet $L$ functions, as 
these functions may be written as a linear combination of Hurwitz zeta functions.  
For instance, for $\chi$ a character modulo $m$ and Re $s > 1$ we have
$$L(s,\chi) = \sum_{k=1}^\infty {{\chi(k)} \over k^s} ={1 \over m^s}\sum_{k=1}^m 
\chi(k) \zeta\left(s,{k \over m}\right).  \eqno(56)$$
Conversely, this relation may be inverted, yielding the Hurwitz zeta function
$\zeta(s,a)$ at rational values of $a$ in terms of a linear combination of $L$
functions. 
When $\chi$ is a non principal character in Eq. (56) convergence obtains for
Re $s>0$.

Now a suitable definition for generalized Stieltjes constants is (cf. \cite{lehmer,knopf})
$$\gamma_k(m)=\lim_{N \to \infty}\left (\sum_{n=2}^N {{\chi(n)} \over {n}}\ln^k n-
{{\ln^{k+1} {N}} \over {m(k+1)}}\right ). \eqno(57)$$


We first note the following from the first equality in (56).  We have
$$\sum_{n=2}^\infty {{(-1)^n} \over n}(-1)^j L^{(j)}(n,\chi)
=\sum_{k=2}^\infty \chi(k)\ln^j k\sum_{n=2}^\infty {{(-1)^n} \over {nk^n}}$$
$$=\sum_{k=2}^\infty \chi(k)\ln^j k\left[{1 \over k} -\ln\left({{k+1} \over k}\right)
\right]$$
$$=\lim_{x \to \infty}\left \{\sum_{k=2}^x \chi(k){{\ln^j k} \over k}+\sum_{k=2}^x
\chi(k)[\ln k-\ln(k+1)]\ln^j k\right \}.  \eqno(58)$$
Putting
$$A(x)=\sum_{k=2}^x \chi(k)[\ln k-\ln(k+1)], \eqno(59)$$
and applying summation by parts we have
$$\sum_{k=2}^x \chi(k)[\ln k-\ln(k+1)]\ln^j k=A(x)\ln^j x-j\int_1^x A(t)
{{\ln^{j-1} t} \over t}dt.  \eqno(60)$$

Alternatively, we now proceed in terms of Hurwitz zeta functions and write from
the product rule and Eq. (56)
$$(-1)^j L^{(j)}(s,\chi)=\sum_{\ell=0}^j (-1)^\ell {j \choose \ell} {{\ln^{j-\ell}m}
\over m^s} \sum_{k=1}^m \chi(k) \zeta^{(\ell)} \left(s,{k \over m}\right).  
\eqno(61)$$
We then have
$$\sum_{n=2}^\infty {{(-1)^n} \over n}(-1)^j L^{(j)}(n,\chi)=\sum_{n=2}^\infty {{(-1)^n} \over n}\sum_{\ell=0}^j (-1)^\ell {j \choose \ell} {{\ln^{j-\ell}m}
\over m^n} \sum_{k=1}^m \chi(k) \zeta^{(\ell)} \left(n,{k \over m}\right).$$  
$$=\sum_{n=2}^\infty {{(-1)^n} \over n}\sum_{\ell=0}^j {j \choose \ell} {{\ln^{j-\ell}m} \over m^n} \sum_{k=1}^m \chi(k)\sum_{q=0}^\infty {{\ln^\ell(q+k/m)} 
\over {(q+k/m)^n}}$$
$$=\sum_{q=0}^\infty \sum_{\ell=0}^j {j \choose \ell} {{\ln^{j-\ell}m} \over m^n} \sum_{k=1}^m \chi(k)\ln^\ell(q+k/m)\left[{1 \over {mq+k}}-\ln\left({{mq+k+1} \over
{mq+k}}\right)\right].  \eqno(62)$$
The sum on $q$ here is
$$\sum_{q=0}^\infty \ln^\ell(q+k/m)\left[{1 \over {mq+k}}-\ln\left({{mq+k+1} \over
{mq+k}}\right)\right]$$
$$=\lim_{x \to \infty}\left \{\sum_{q=0}^x {{\ln^\ell(q+k/m)} \over {mq+k}} +\sum_{q=0}^x [\ln (mq+k)-\ln(mq+k+1)]\ln^\ell (q+k/m)\right \}.  \eqno(63)$$
Summing by parts, we have
$$\sum_{q=0}^x [\ln (mq+k)-\ln(mq+k+1)]\ln^\ell (q+k/m)=[\ln k-\ln(k+1)]\ln^\ell (k/m)$$
$$+B(x)\ln^\ell (x+k/m)-\ell\int_0^xB(t){{\ln^{\ell-1} (t+k/m)} \over {t+k/m}}dt,  
\eqno(64)$$
where we have put
$$B(x) = \sum_{q=1}^x [\ln(mq+k)-\ln(mq+k+1)].  \eqno(65)$$
The integral term of Eq. (64) gives an extension of the integral of Eq. (26) when
$m>1$.


From Proposition 1(b) we obtain
{\newline \bf Corollary 1}.  We have for Re $a>0$ and Re $b>0$
$$\ln\left[{{\Gamma(b)} \over {\Gamma(a)}}\right]=b\ln^2 b-a\ln^2a+\int_0^\infty
[1+\ln(a+t)]\ln\left({{[t]+a+1} \over {t+a+1}}\right)dt-\int_0^\infty
[1+\ln(b+t)]\ln\left({{[t]+b+1} \over {t+b+1}}\right)dt$$
$$+\sum_{k=0}^\infty \ln(k+a)\left[1-(k+a)\ln\left({{k+a+1}\over {k+a}}\right)\right]
-\sum_{k=0}^\infty \ln(k+b)\left[1-(k+b)\ln\left({{k+b+1}\over {k+b}}\right)\right]$$
$$+b+\ln b[-b+(1+b)\ln(b+1)]+\mbox{Li}_2(-b)
-a+\ln a[a-(1+a)\ln(a+1)]-\mbox{Li}_2(-a).  \eqno(66)$$

In contrast, a standard sum representation is \cite{grad} (p. 936)
$$\ln\left[{{\Gamma(b)} \over {\Gamma(a)}}\right]=\ln\left({a \over b}\right)
+\sum_{k=1}^\infty \left[(b-a)\ln\left(1+{1 \over k}\right)+\ln\left({{k+a} 
\over {k+b}}\right)\right].   \eqno(67)$$

Proof.  We use the summatory relation \cite{coffeystdiffs} (Proposition 1)
$$\sum_{n=0}^\infty {1 \over {n!}}[\gamma_{n+1}(a)-\gamma_{n+1}(b)]=\ln\left[
{{\Gamma(b)} \over {\Gamma(a)}}\right], ~~~~\mbox{Re} ~a>0, ~~\mbox{Re} ~b>0. 
\eqno(68)$$
We write
$$(-1)^j\sum_{m=2}^\infty {{(-1)^m} \over m}\left[\zeta^{(j)}(m,a)-\zeta^{(j)}(m,b)
\right]=\sum_{k=0}^\infty \ln^j(k+a)\left[{1 \over {k+a}}-\ln\left({{k+a+1} \over
{k+a}}\right)\right]$$
$$-\sum_{k=0}^\infty \ln^j(k+b)\left[{1 \over {k+b}}-\ln\left({{k+b+1} \over
{k+b}}\right)\right].  \eqno(69)$$
Putting $j=n+1$ and summing $\sum_{n=0}^\infty (1/n!)$ gives the terms of line $2$
of the Corollary.  The integral terms of the Corollary result from using the
definition (5) of $d_{n+1}(y)$.

For the polylogarithm terms we have
$$\sum_{n=0}^\infty {{(n+1)} \over {n!}}\sum_{\ell=0}^n (-1)^\ell \ell!
{n \choose \ell} \ln^{n-\ell} a ~\mbox{Li}_{\ell+2}(-a)$$
$$=\sum_{\ell=0}^\infty (-1)^\ell \ell! \mbox{Li}_{\ell+2}(-a)\sum_{n=\ell}^\infty
{{(n+1)} \over {n!}}\sum_{\ell=0}^n {n \choose \ell}\ln^{n-\ell} a $$
$$=a \sum_{\ell=0}^\infty (-1)^\ell (1+\ln a +\ell)\mbox{Li}_{\ell+2}(-a)$$
$$=a\sum_{j=1}^\infty (-a)^j \left[{1 \over {(j+1)^2}}+{{\ln a} \over {j(j+1)}}
\right]$$
$$=-a+\ln a[-a+(1+a)\ln(a+1)]+\mbox{Li}_2(-a).  \eqno(70)$$
Combining terms completes the Corollary.

Akin to Eq. (24) we have from Eqs. (5) and (6) the weak upper asymptotic bounds
$$|c_j| = O[(j-1)!], ~~~~~~|d_j(a)|=O[\Gamma(j,\ln a)], ~~j \to \infty.  \eqno(71)$$ 
These bounds miss cancellation, so are large over estimates.

Finally in closing this section we write a summation form for the constants 
$d_j(a)$.  From Eq. (5) and the use of Lemma 1 we have
$$d_j(a)=\sum_{k=0}^\infty \int_{k+a}^{k+a+1} {{\ln^{j-1} u} \over u} \left[\ln(k+a+1)-\ln(u+1) \right]du$$
$$={1 \over j}\sum_{k=0}^\infty \left[\ln^j(k+a+1)-\ln^j(k+a) \right]\ln(k+a+1)$$
$$-\sum_{k=0}^\infty\sum_{\ell=0}^{j-1}(-1)^\ell \ell!{{j-1} \choose \ell}\left[
\ln^{j-\ell-1}(k+a+1) \mbox{Li}_{\ell+2}(-k-a-1)-\ln^{j-\ell-1}(k+a) \mbox{Li}_{\ell+2}(-k-a)\right].  \eqno(72)$$

\medskip
\centerline{\bf Other expressions for $\gamma_0(a)$}
\medskip

Recalling that $\gamma_0(a)=-\psi(a)$, we have
{\newline \bf Proposition 2}.  We have (a)
$$\gamma_0(a)=-{1 \over a}+\ln(a+1)-\sum_{m=2}^\infty {{(-1)^m} \over m}
\left[\zeta(m,a)-{1 \over a^m}\right ],  \eqno(73)$$
(b)
$$\sum_{m=2}^\infty{{(-1)^m} \over {m+1}}\zeta(m,a)=-a\ln a+\ln \Gamma(a+1)
-{1 \over 2} \psi(a)-{1 \over 2}\ln(2 \pi)+a,  \eqno(74)$$
and for $|z|<|a|$ (c)
$$\sum_{m=2}^\infty{{z^m} \over m}\left[\zeta(m,a)-{1 \over a^m}\right]
={z \over a}+\ln\left(1-{z \over a}\right)-\ln \Gamma(a-z)+\ln\Gamma(a)
-z\psi(a).  \eqno(75)$$

Proof.  For part (a), we form
$$\sum_{m=2}^\infty {{(-1)^m} \over m}\left[\zeta(m,a)-{1 \over a^m}\right ]
=\sum_{m=2}^\infty {{(-1)^m} \over m} \sum_{k=1}^\infty {1 \over {(k+a)^m}}$$
$$=\sum_{k=1}^\infty \left[{1 \over {k+a}}-\ln(k+a+1)+\ln(k+a)\right]$$
$$=\ln(1+a)+\lim_{x \to \infty}\left[\sum_{k=1}^x {1 \over {k+a}}
-\ln([x]+a+1)\right] \eqno(76)$$
$$=\ln(1+a)-\psi(a+1)+\lim_{x \to \infty}\left[\psi([x]+a+1)-\ln([x]+a+1)
\right]$$
$$=\ln(1+a)-\psi(a+1)+\lim_{x \to \infty} O\left({1 \over {[x]+a+1}}\right)$$
$$=\ln(1+a)-\psi(a)-{1 \over a}.  \eqno(77)$$
In the last step we have used the functional equation of the digamma function.

An alternative method of proving part (a) is by applying the integral 
representation (35).  Then we have
$$\sum_{m=2}^\infty {{(-1)^m} \over m}\left[\zeta(m,a)-{1 \over a^m}\right ]
=\sum_{m=2}^\infty {{(-1)^m} \over m}\left[{1 \over {(m-1)!}}\int_0^\infty
{{t^{m-1} e^{-(a-1)t}} \over {e^t-1}}dt - {1 \over a^m}\right]$$
$$=-{1 \over a}+\ln\left(1 + {1 \over a}\right)-\sum_{m=2}^\infty {{(-1)^m} \over
{m!}} \int_0^\infty {{t^{m-1} e^{-(a-1)t}} \over {e^t-1}}dt$$
$$=-{1 \over a}+\ln\left(1 + {1 \over a}\right)-\int_0^\infty {{e^{-(a-1)t}} \over
{e^t-1}}{1 \over t}(e^{-t}+t-1)dt$$
$$=-{1 \over a}+\ln\left(1 + {1 \over a}\right)-\ln a + \psi(a)$$
$$=-{1 \over a}+\ln(a+1)+\psi(a).  \eqno(78)$$
The absolute convergence of the integral representation (35) justifies
the interchange of summation and integration above.

For part (b), we have
$$\sum_{m=2}^\infty {{(-1)^m} \over {m+1}}\zeta(m,a)=\sum_{m=2}^\infty {{(-1)^m} 
\over {(m+1)}} \sum_{k=0}^\infty {1 \over {(k+a)^m}}$$
$$=\sum_{k=0}^\infty\left\{(k+a)[\ln(k+a+1)-\ln(k+a)]+{1 \over 2}{1 \over {(k+a)}}
-1 \right\}$$
$$=\lim_{x \to \infty}\sum_{k=0}^x \left\{(k+a+1)\ln(k+a+1)-(k+a)\ln(k+a)
-\ln(k+a+1)+{1 \over 2}{1 \over {(k+a)}}-1 \right\} \eqno(79)$$
$$=\lim_{x \to \infty}\left\{([x]+a+1)\ln([x]+a+1)-a \ln a+\ln \Gamma(a+1)
-\ln\Gamma([x]+a+2) \right.$$
$$\left.+{1 \over 2}[\psi([x]+a+1)-\psi(a)]-([x]+1) \right \}.  \eqno(80)$$
We then apply Stirling's formula for $\ln \Gamma(z)$ and part (b) follows.

For part (c), we proceed as in the second proof of part (a).  We have
$$\sum_{m=2}^\infty {{z^m} \over m}\left[\zeta(m,a)-{1 \over a^m}\right ]
=\sum_{m=2}^\infty {{z^m} \over m}\left[{1 \over {(m-1)!}}\int_0^\infty
{{t^{m-1} e^{-(a-1)t}} \over {e^t-1}}dt - {1 \over a^m}\right]$$
$$={z \over a}+\ln\left(1 - {z \over a}\right)-\int_0^\infty {{e^{-(a-1)t}}
\over {(e^t-1)}}{1 \over t}(e^{zt}-zt-1)dt.  \eqno(81)$$
We use \cite{grad} (p. 939)
$$\ln \Gamma(a-z)-\ln\Gamma(a)=\int_0^\infty \left[{{e^{-(a-z)t}-e^{-at}} \over
{1-e^{-t}}}-z e^{-t}\right]{{dt} \over t}, \eqno(82)$$
along with an integral representation of $\gamma$ plus the digamma function
\cite{grad} (p. 943) and of the Euler constant \cite{grad} (p. 946) in the form
$$\gamma=\int_0^\infty \left[{e^{-t} \over {1-e^{-t}}}-{e^{-t} \over t}\right]
dt.  \eqno(83)$$
Proposition 2 is finished.

Remark.  Part (a) of this Proposition gives the $a \neq 1$ extension of Ref. 
\cite{verma}.  Part (b) gives the $a \neq 1$ extension of Refs. \cite{singh}
and \cite{suryan}.


\pagebreak
\centerline{\bf Generalizations of sum of Hurwitz zeta function values}
\medskip

Here we generalize the sums of the previous section, showing 
{\newline \bf Proposition 3}.  For integers $j \geq 1$ we have
$$\sum_{m=2}^\infty {{(-1)^m} \over {m+j}}\zeta(m,a)=(-1)^j \lim_{x \to \infty}
\sum_{k=0}^x \left\{(k+a)^{j-1}-(k+a)^j \ln\left(1+{1 \over {k+a}}\right)\right.$$
$$\left.-(k+a)^j \sum_{m=2}^{j+1} {{(-1)^m} \over m} {1 \over {(k+a)^m}} \right \}.
\eqno(84)$$

The proof of this Proposition depends upon
{\newline \bf Lemma 4}.  For integers $j \geq 0$ and $k \in C$ let
$$M_j(k) \equiv \sum_{m=2}^\infty {{(-1)^m} \over {(m+j)}}{1 \over {(k+a)^m}}.
\eqno(85)$$
Then we have the expressions
$$M_j(k)={1 \over {(j+2)}}{1 \over {(k+a)^2}} ~_2F_1\left(1,j+2;j+3;-{1 \over
{k+a}}\right) \eqno(86)$$
$$=(-1)^j (k+a)^jM_0(k)-(-1)^j(k+a)^j\sum_{m=2}^{j+1}{{(-1)^m}
\over m} {1 \over {(k+a)^m}}, \eqno(87)$$
where $_2F_1$ is the Gauss hypergeometric function and 
$$M_0(k)={1 \over {k+a}}-\ln\left(1+{1 \over {k+a}}\right).  \eqno(88)$$

For the proof of Eq. (86), let $(a)_n=\Gamma(a+n)/\Gamma(a)$ be the
Pochhammer symbol.  By noting that $(j+2)_m/(j+3)_m=(j+2)/(j+m+2)$ we have
$$M_j(k)=\sum_{m=0}^\infty {{(-1)^m} \over {(m+j+2)}}{1 \over {(k+a)^{m+2}}}
={1 \over {(j+2)}}{1 \over {(k+a)^2}}\sum_{m=0}^\infty {{(1)_m} \over {m!}}
{{(j+2)_m} \over {(j+3)_m}} \left({{-1} \over {k+a}}\right)^m.  \eqno(89)$$
By the series definition of the function $_2F_1$, Eq. (86) holds.
In particular, $M_0(k)$ is a known special case.

For Eq. (87), we observe by shifting the summation index that
$$M_{j+1}(k)=-(k+a)M_j(k) + {1 \over {(j+2)}}{1 \over {(k+a)}}.  \eqno(90)$$
In addition to this recursion relation we have
$$M_j(k)=(-1)^j(k+a)^j \sum_{m=j+2}^\infty {{(-1)^m} \over m}{1 \over {(k+a)^m}},
\eqno(91)$$
and Eq. (87) follows.

For the proof of Proposition 3, we write
$$\sum_{m=2}^\infty {{(-1)^m} \over {m+j}}\zeta(m,a)=\sum_{k=0}^\infty M_j(k),
\eqno(92)$$
and then apply Lemma 4.

The manner in which the sums of terms rational in $k+a$ in Proposition 3 may be
evaluated in terms of polygamma functions $\psi^{(j)}$ is given by
{\newline \bf Lemma 5}.  We have (a)
$$\sum_{k=0}^x (k+a)^{j-m}={{(-1)^{m-j}} \over {(m-j-1)!}}\left[\psi^{(m-j-1)}
(a)-\psi^{(m-j-1)}([x]+a+1)\right],  \eqno(93)$$
and (b) the asymptotic value of the second term on the right side of Eq. (93) is
given by
$$-{{(-1)^{m-j}} \over {(m-j-1)!}}\psi^{(m-j-1)}([x]+a+1)
={1 \over {(m-j-1)}}{1 \over {([x]+a+1)^{m-j-1}}}+O\left({1 \over {([x]+a+1)^{m-j}}}
\right),$$
$$~~~~~~~~~~~~~~~~~~~~~~~~~~~~~~~~~~~~~~~~~~~~~~~~~~~~~~~~~~~~~~~~~~~~~~~~x \to \infty. \eqno(94)$$

For the proof of part (a) of this Lemma, we note that
$$\sum_{k=0}^x (k+a)^{j-m}=\zeta(m-j,a)-\zeta(m-j,[x]+a+1).  \eqno(95)$$
We then use the relation 
$$\zeta(n+1,x)={{(-1)^{n+1}} \over {n!}} \psi^{(n)}(x), ~~~~~~~n \geq 1.
\eqno(96)$$
For part (b) we apply the known asymptotic form of the polygamma functions
\cite{nbs}.

Remarks.  The asymptotic evaluation of terms of the form
$(k+a)^j[\ln(k+a+1)-\ln(k+a)]$ in Proposition 3 may be performed by way of
Euler-Maclaurin summation.  So at $j=1$ one way to write the result is in
terms of the Glaisher-Kinkelin constant.  Similarly, at $j=2$ and $j=3$ the
results may be written in terms of constants $B$ and $C$ that have previously
appeared in the literature (e.g., \cite{sri}, pp. 36 and 37). 

Examination of Eq. (95) shows that $j$ there need not be an integer.
We may therefore differentiate it with respect to $j$ and find
$$\sum_{k=0}^x (k+a)^{j-m}\ln(k+a)=\zeta'(m-j,[x]+a+1)-\zeta'(m-j,a),
\eqno(97)$$
where $'$ denotes differentiation with respect to the first argument.
Also for use in connection with Proposition 3 we have by the integral
representation (35) and Euler-Maclaurin summation
$\zeta(s,a+x)=x^{1-s}/(s-1)+O(x^{-s})$ as $x \to \infty$, giving
$$\zeta'(s,a+x)={x^{1-s} \over {s-1}}\ln x - {x^{1-s} \over {(s-1)^2}}
+O(x^{-s} \ln x), ~~~~~~~~~~ x \to \infty.  \eqno(98)$$

Additionally we may further generalize Proposition 3 to 
{\newline \bf Proposition 4}.  We have for Re $a>0$, $j >-1$, and $|t| < 1$ (a)
$$\sum_{k=2}^\infty (-1)^k \zeta(k,a) {t^{k+j} \over {k+j}}=\int_0^t
x^j[\psi(x+a)-\psi(a)]dx, \eqno(99)$$
and (b)
$$(-1)^m\sum_{k=2}^\infty (-1)^k (k)_m\zeta(k+m,a) {t^{k+j} \over {k+j}}
=\int_0^tx^j[\psi^{(m)}(x+a)-\psi^{(m)}(a)]dx, \eqno(100)$$
or, equivalently, (c)
$$\sum_{k=2}^\infty (-1)^k (k)_m\zeta(k+m,a) {t^{k+j} \over {k+j}}
=m!\int_0^tx^j[\zeta(m+1,x+a)-\zeta(m+1,a)]dx. \eqno(101)$$

Part (c) follows from part (b) by the use of relation (96).
For the proof of part (a) we start with the identity
$$\sum_{k=2}^\infty (-1)^k \zeta(k,a) x^{k-1}=\psi(x+a)-\psi(a).  \eqno(102)$$
We then multiply both sides of this equation by $x^j$ and integrate.
The condition on $j$ is imposed due to the fact that $\psi(x)=-1/x-\gamma
+O(x)$ as $x \to 0$.
For part (b) we repeatedly differentiate Eq. (75) with respect to $a$, use
$${\partial^m \over {\partial a^m}}\zeta(k,a)=(-1)^m (k)_m \zeta(k+m,a),
\eqno(103)$$
and the Proposition is completed.

Remark.  Proposition 4 may be extended by multiplying Eq. (102) by other
functions and integrating over other intervals.  For instance we have
$$\sum_{k=2}^\infty (-1)^k \zeta(k,a)\int_0^t x^{k+\beta-1}(1-x)^\gamma
(1-zx)^{-\alpha}dx=\int_0^t x^\beta (1-x)^\gamma (1-zx)^{-\alpha}[\psi(x+a)
-\psi(a)]dx.  \eqno(104)$$
Once again such a result may be differentiated with respect to $a$
and Eqs. (103) and (96) used in order to obtain certain integrals of
differences of Hurwitz zeta functions.
Moreover, Eq. (100) could again be multiplied by say $t^n$ and integrated.
For special values of $t$ including $1$ explicit results are obtainable
by using various representations of the digamma function.  

By using various integral representations of the Hurwitz zeta function we
may determine evaluations for the integrals of Proposition 4.  We illustrate
this in relation to Eq. (92).  If we put for integers $j \geq 0$, $k \in C$,
and $|t|<1$,
$$M_j(k,t,a) \equiv \sum_{m=2}^\infty {{(-1)^m} \over {(m+j)}}{t^m \over {(k+a)^m}},
\eqno(105)$$
then we have 
$$M_j(k,t,a)={1 \over {(j+2)}}{t^2 \over {(k+a)^2}} ~_2F_1\left(1,j+2;j+3;-{t \over
{k+a}}\right) \eqno(106)$$
$$=(-1)^j t^{-j}(k+a)^j\left[M_0(k,t,a)-\sum_{m=2}^{j+1}{{(-1)^m t^m}
\over {m(k+a)^m}}\right], \eqno(107)$$
and 
$$M_0(k,t,a)={t \over {k+a}}-\ln\left(1+{t \over {k+a}}\right).  \eqno(108)$$
These expressions follow in like manner to Lemma 4.  We then have
{\newline \bf Proposition 5}.  For Re $a>0$, integers $j \geq 0$, and $|t| < 1$
we have 
$$\sum_{k=2}^\infty (-1)^k \zeta(k,a) {t^{k} \over {k+j}}=\left({1 \over 2}-
{a \over {(j+1)}}\right)M_j(0,t,a)+{t \over {j+1}}\ln(1+t/a)$$
$$-\int_0^\infty\left[{t^2 \over
{(x+a)^2}}{1 \over {(x+t+a)}}-jM_j(x,t,a)\right]P_1(x)dx \eqno(109)$$
$$=\left({1 \over 2}-{a \over {(j+1)}}\right)M_j(0,t,a)+{t \over {j+1}}\ln(1+t/a)
-2\int_0^\infty \sum_{k=2}^\infty {{\sin(k \tan^{-1} y/a)} \over {(a^2+y^2)^{k/2}}}
{{dy} \over {(e^{2\pi y}-1)}}  \eqno(110)$$
$$=\left({1 \over 2}-{a \over {(j+1)}}\right)M_j(0,t,a)+{t \over {j+1}}\ln(1+t/a)$$
$$+\int_0^\infty \left({1 \over {1-e^{-w}}}-{1 \over w}-{1 \over 2}\right){e^{-aw}
\over w} M_j(0,t,1/w)dw.  \eqno(111)$$

In the proof of Eq. (109) we use the representation
$$\zeta(s,a)={a^{-s} \over 2} +{a^{1-s} \over {s-1}}-s\int_0^\infty {{P_1(x)} \over {(x+a)^{s+1}}}dx, ~~~~~~~\mbox{Re} ~s > -1,  \eqno(112)$$
that is based upon Euler-Maclaurin summation.  For the proof of Eq. (93) we use
Hermite's formula for the Hurwitz zeta function (e.g., \cite{watson} p. 270),
$$\zeta(s,a)={a^{-s} \over 2}+{a^{1-s} \over {s-1}}+2\int_0^\infty {{\sin(s\tan^{-1}y/a)} \over {(y^2+a^2)^{s/2}}}{{dy} \over {(e^{2\pi y}-1)}}.
\eqno(113)$$

Similarly, we may use a different integral representation for $\zeta(k,a)$ to
obtain Eq. (111).  We however describe a second method based upon the relation
(96), whereby we obtain
$$\sum_{k=2}^\infty (-1)^k \zeta(k,a) {t^{k} \over {k+j}}=\sum_{k=2}^\infty
{t^k \over {(k+j)}}{1 \over {(k-1)!}}\psi^{(k-1)}(a).  \eqno(114)$$
We make use of a classical Binet formula for the digamma function \cite{sri}.
Upon repeated differentiation, we find
$$\psi^{(n)}(z)=(-1)^{n-1}{{(n-1)!} \over z^n}-{{(-1)^n n!} \over {2z^{n+1}}}
-(-1)^n \int_0^\infty\left[{1 \over {1-e^{-t}}}-{1 \over t}-{1 \over 2}
\right]t^n e^{-tz} dt, ~~~~\mbox{Re}~z > 0.  \eqno(115)$$
Based upon this expression, inserting $\psi^{(k-1)}(a)$ into Eq. (114)
and again interchanging summation and integration we determine Eq. (111).
The interchange is justified by the absolute convergence of the integral of
Eq. (115) and Proposition 5 is completed.

Remarks.  It is possible to re-express the summation on the right side of Eq. 
(110).  It is also possible to insert Binet's formula for $\psi$ and $\psi^{(m)}$
into Eqs. (99) and (100) and thereby obtain other expressions.  This too we omit.

\medskip
\centerline{\bf An extension and brief discussion}
\medskip

In principle, a result of the shape of Proposition 1 holds for all 
convergent Dirichlet series.  Rather than describe this generally, we
illustrate this for a Dirichlet series of particular interest.  Namely,
for 
$${{\zeta'} \over {\zeta}}(s)=-\sum_{k=2}^\infty {{\Lambda(k)} \over k^s},
~~~~~~~~~~\mbox{Re} ~s >1, \eqno(116)$$
where $\Lambda$ is the von Mangoldt function \cite{ivic,titch}.  Like
$\zeta$ itself, the logarithmic derivative (116) has a simple pole at $s=1$
and the Laurent expansion analogous to Eq. (1) is 
$${{\zeta'(s)} \over {\zeta(s)}}=-{1 \over {s-1}}-\sum_{p=0}^\infty 
\eta_p (s-1)^p, ~~~~~~|s-1| < 3, \eqno(117)$$ 
wherein the constants $\eta_j$ can be written as
$$\eta_k={(-1)^k \over {k!}} \lim_{N \to \infty}\left (\sum_{m=1}^N {1 \over 
m} \Lambda(m)\ln^k m - {{\ln^{k+1} N} \over {k+1}}\right ). \eqno(118)$$
The $\eta_j$ coefficients are related to the Stieltjes constants $\gamma_k$
by a simple recurrence relation (e.g., \cite{coffeyjcam04}, Appendix). 

We now have
\newline{\bf Proposition 6}.  For integers $j \geq 0$ we have
$$\eta_j = -{1 \over {j!}}\sum_{m=2}^\infty {{(-1)^m} \over m}\left ({{\zeta'}
\over \zeta}\right)^{(j)}(m)+{{(-1)^j} \over {j!}}\lim_{N \to \infty}
\left[\sum_{k=2}^N \Lambda(k) \ln^j k \ln\left({{k+1} \over k}\right)-
{{\ln^{j+1} N} \over {j+1}}\right], \eqno(119)$$
and
\newline{\bf Corollary 2}.  We have
$$\eta_0=-\gamma = -\sum_{m=2}^\infty {{(-1)^m} \over m}{{\zeta'}
\over \zeta}(m)+\lim_{N \to \infty}\left[\sum_{k=2}^N \Lambda(k)
\ln\left({{k+1} \over k}\right)-\ln N \right].   \eqno(120)$$

On the right side of Eq. (119), the dominant asymptotic order $\ln^{j+1}
N/(j+1)$ is cancelled by the sum, leading to a new set of constants
analogous to the second and third terms on the right side of Eq. (7),
whose explicit determination remains open.  Again partial summation may
be applied to the right side of Eq. (119) and alternative forms written.
As we do not wish to invoke the Riemann hypothesis, we do not pursue this
here.

An approximate value of the first contribution on the right side of Eq. (120)
is $0.2419488514703058$, enabling a numerical approximation for the constant
coming from the limit term.  
In comparison, the term $\zeta'(2)/[2\zeta(2)]=(\gamma-\ln 2+\ln \pi+12\ln A)/2
\simeq -0.2849804965472664$ where $A$ is Glaisher's constant.

For the proof of Proposition 6 we form
$${1 \over {j!}}\sum_{m=2}^\infty {{(-1)^m} \over m}\left({{\zeta'} \over \zeta}\right)^{(j)}(m)=-{{(-1)^j} \over {j!}}\sum_{k=2}^\infty \Lambda(k)\ln^j 
k\left[{1 \over k} -\ln\left({{k+1} \over k}\right)
\right].  \eqno(121)$$
We then add the expression (118) and Eq. (119) follows.  

The importance of a new representation of the $\eta_j$ coefficients lies
especially with new representation and estimation of the critical
alternating binomial sum
$$S_2(n) \equiv -\sum_{m=1}^n {n \choose m} \eta_{m-1}
=\sum_{m=1}^n (-1)^m {n \choose m} |\eta_{m-1}|.  \eqno(122)$$
The linearity or sublinearity of $S_2(n) = S_\gamma(n) + S_\Lambda(n)$ in $n$,
wherein the oscillating sum $S_\Lambda(n)$ involves logarithms and values
of the von Mangoldt function \cite{inprep}, more than suffices to
verify the Riemann hypothesis under the Li criterion.  Indeed, the 
Riemann hypothesis fails only if $S_\Lambda(n)$ and hence $S_2(n)$
becomes exponentially large in $n$ and negative.




\pagebreak


\begin{thebibliography}{99}
\bibitem{nbs}M. Abramowitz and I. A. Stegun,
{Handbook of Mathematical Functions, Washington, National Bureau of Standards
(1964).}
\bibitem{coffeyjmaa}M. W. Coffey,
{New results on the Stieltjes constants:  Asymptotic and exact evaluation, 
J. Math. Anal. Appl. {\bf 317}, 603-612 (2006); arXiv:math-ph/0506061.}
\bibitem{coffeyprsa}M. W. Coffey,
{New summation relations for the Stieltjes constants, Proc. Royal Soc. A
{\bf 462}, 2563-2573 (2006).}
\bibitem{inprep}M. W. Coffey,
{The Stieltjes constants, their relation to the $\eta_j$ coefficients,
and representation of the Hurwitz zeta function, arxiv/math.ph/0706343 (2007).}
\bibitem{coffeympag}M. W. Coffey,
{Towards verification of the Riemann hypothesis, Math. Phys., Analysis and
Geometry {\bf 8}, 211-255 (2005).}
\bibitem{coffey08}M. W. Coffey,
{New results on power series expansions of the Riemann xi function and the
Li/Keiper constants, Proc. Royal Soc. A {\bf 464}, 211 (2008).}
\bibitem{coffeyjcam04}M. W. Coffey,
{Relations and positivity results for derivatives of the Riemann $\xi$
function, J. Comput. Appl. Math., {\bf 166}, 525-534 (2004).}
\bibitem{coffeystdiffs}M. W. Coffey,
{On representations and differences of Stieltjes coefficients, and other relations,
arXiv/math-ph/0809.3277 (2008).}
\bibitem{grad}I. S. Gradshteyn and I. M. Ryzhik,
{Table of Integrals, Series, and Products, Academic Press, New York (1980).}
\bibitem{hansen}E. R. Hansen and M. L. Patrick,
{Some relations and values for the generalized Riemann zeta function,
Math. Comp. {\bf 16}, 265-274 (1962).}
\bibitem{ivic}A. Ivi\'{c}, 
{The Riemann Zeta-Function, Wiley (1985).}
\bibitem{knopf}J. Knopmacher,
{Generalized Euler constants, Proc. Edinburgh Math. Soc. {\bf 21}, 25-32 (1978).}
\bibitem{lehmer}D. H. Lehmer,
{Euler constants for arithmetic progressions, Acta Arith. {\bf 27}, 125-142 (1975).}
\bibitem{liang}J. J. Y. Liang and J. Todd,
{The Stieltjes constants, J. Res. Natl. Bur. Stand. {\bf 768}, 161-178 (1972).}
\bibitem{singh}R. J. Singh and D. P. Verma,
{Some series involving Riemann zeta function, Yokohama Math. J. {\bf 31},
1-4 (1983).}
\bibitem{sri}H. M. Srivastava and J. Choi,
{Series associated with the zeta and related functions, Kluwer (2001).}
\bibitem{stieltjes}T. J. Stieltjes,
{Correspondance d'Hermite et de Stieltjes, Volumes 1 and 2, Gauthier-Villars,
Paris (1905).}
\bibitem{suryan}D. Suryanarayana,
{Sums of Riemann zeta function, Math. Student, {\bf 42}, 141-143 (1974).}
\bibitem{titch}E. C. Titchmarsh,
{The Theory of the Riemann Zeta-Function, 2nd ed., Oxford University
Press, Oxford (1986).}
\bibitem{verma}D. P. Verma,
{A note on Euler's constant, Math. Student {\bf 29}, 140-141 (1961).}
\bibitem{watson}E. T. Whittaker and G. N. Watson,
{A course of modern analysis, Cambridge University Press (1973).}
\bibitem{wilton2}J. R. Wilton, 
{A note on the coefficients in the expansion of $\zeta(s,x)$ in powers of 
$s-1$, Quart. J. Pure Appl. Math. {\bf 50}, 329-332 (1927).}
\end{thebibliography}
\end{document}